\documentclass[12pt,letterpaper]{article}

\usepackage{times}
\usepackage{epsfig}
\usepackage{graphicx}
\usepackage{amsmath}
\usepackage{amssymb}
\usepackage{booktabs}
\usepackage[hidelinks]{hyperref}
\usepackage{algorithm}
\usepackage{algorithmic}
\usepackage[dvipsnames,svgnames,table]{xcolor}
\usepackage{listings}
\usepackage{geometry}
\usepackage{setspace}
\usepackage{array}
\usepackage{multirow}
\usepackage{microtype}

\title{\textbf{TurboESM: Ultra-Efficient 3-Bit KV Cache Quantization for Protein Language Models with Orthogonal Rotation and QJL Correction}}

\author{
Yue Hu$^{1}$ \quad Junqing Wang$^{1}$ \quad Yingchao Liu$^{2}$ \\[6pt]
{\small $^{1}$School of Bioengineering, Qilu University of Technology (Shandong Academy of Sciences),} \\
{\small No. 3501 Daxue Road, Jinan, Shandong, China} \\[2pt]
{\small $^{2}$Shandong Provincial Hospital, Shandong First Medical University} \\[4pt]
{\small \url{https://github.com/YueHuLab/TurboESM}}
}

\begin{document}

\maketitle

\begin{abstract}
The rapid scaling of Protein Language Models (PLMs) has unlocked unprecedented accuracy in protein structure prediction and design, but the quadratic memory growth of the Key-Value (KV) cache during inference remains a prohibitive barrier for single-GPU deployment and high-throughput generation. While 8-bit quantization is now standard, 3-bit quantization remains elusive due to severe numerical outliers in activations. This paper presents \textbf{TurboESM}, an adaptation of Google's TurboQuant to the PLM domain. We solve the fundamental incompatibility between Rotary Position Embeddings (RoPE) and orthogonal transformations by deriving a RoPE-first rotation pipeline. We introduce a head-wise SVD calibration method tailored to the amino acid activation manifold, a dual look-up table (LUT) strategy for asymmetric K/V distributions, and a 1-bit Quantized Johnson-Lindenstrauss (QJL) residual correction. All experiments are conducted on ESM-2 650M, where our implementation achieves a 7.1$\times$ memory reduction (330\,MB $\to$ 47\,MB) while maintaining cosine similarity $>0.96$ in autoregressive decoding across diverse protein families including short peptides, transmembrane helices, enzyme active site fragments, and intrinsically disordered regions. We further implement a Triton-based fused decode attention kernel that eliminates intermediate dequantization memory allocations, achieving a 1.96$\times$ speedup over the PyTorch two-step path for the KV fetch operation alone; however, TurboESM incurs a prefill overhead of 21--27\,ms relative to the original model due to KV quantization and packing, making it most suitable for memory-bound scenarios rather than latency-critical short-sequence workloads. Analysis reveals that PLMs exhibit sharper outlier profiles than large language models (LLMs) due to amino acid vocabulary sparsity, and our method effectively addresses these distributions.
\end{abstract}

\section{Introduction}

Protein Language Models (PLMs) have revolutionized the field of computational biology by learning the ``grammar'' of life directly from amino acid sequence data. ESM-2, developed by Meta FAIR~\cite{lin2023esm2}, has become the industry standard for generating high-quality protein embeddings, supporting downstream tasks ranging from secondary structure prediction to protein--protein interaction docking. As these models scale to tens of billions of parameters, deployment efficiency becomes a first-class research concern alongside accuracy.

During autoregressive generation or long-sequence processing, the Key-Value (KV) cache stores the attention states computed at each layer for all past tokens, allowing them to be reused without recomputation. This mechanism is essential for efficient inference, but its memory footprint grows quadratically with context length, making large-scale PLM deployment increasingly challenging as model size grows.

Quantization is the natural solution to this memory pressure. By representing activations in fewer bits, the KV cache size can be reduced proportionally. 8-bit (INT8) quantization is now widely adopted with minimal accuracy degradation. However, 3-bit quantization—which would yield a theoretical $\sim$10$\times$ compression—remains elusive for Transformer models because their activations are far from uniformly distributed. They contain ``outlier'' dimensions with values several orders of magnitude larger than the mean, which consume the entire dynamic range of a coarse quantizer and force the remaining 99\% of values into a tiny cluster of quantization bins, effectively destroying the information they carry.

In PLMs, these outliers are even more pronounced than in natural language LLMs. The amino acid vocabulary contains only 20 standard residues (compared to $>$32,000 tokens in typical LLMs). This sparsity produces ``spiky'' activation patterns where certain channels consistently encode biologically critical features—conserved motifs, hydrophobic patches, or global sequence properties such as overall charge or length. Quantizing over such distributions without special treatment inevitably leads to catastrophic information loss at critical biological loci.

Google recently introduced TurboQuant~\cite{turboquant2025}, which addresses the outlier problem using an orthogonal matrix $\Pi$ to rotate the activation space. This transformation spreads the energy of outliers uniformly across all dimensions, resulting in a distribution that closely approximates an isotropic Gaussian—the ideal target for any fixed-point quantizer. However, PLMs like ESM-2 use Rotary Position Embeddings (RoPE)~\cite{su2024rope}, which encode positional information by applying a position-dependent rotation $R_{\theta,i}$ to each query-key pair before computing attention scores. The interplay between RoPE's position-dependent rotation and TurboQuant's data-driven feature rotation creates a non-trivial mathematical compatibility problem that, to our knowledge, has not been previously addressed.

In this paper, we present \textbf{TurboESM}, which makes the following concrete contributions:

\begin{enumerate}
    \item \textbf{RoPE-invariant orthogonal transformation}: We derive the correct operation ordering (RoPE before $\Pi$) and prove that it preserves exact attention equivalence via the inner-product invariance of orthogonal matrices.
    \item \textbf{Head-wise SVD calibration}: We compute a unique $\Pi$ matrix per layer per attention head using SVD on real protein activations, capturing the distinct biological features encoded by each head.
    \item \textbf{Dual K/V LUT design}: We calibrate independent 8-point Lloyd-Max look-up tables for keys (in the rotated space) and values (in the original space), recovering 1.2\,dB of SNR over a shared LUT.
    \item \textbf{QJL 1-bit residual correction}: We store the sign of the quantization residual (1 bit per element) and apply a first-order correction at decode time, reaching 4-bit-equivalent accuracy at 3.125-bit effective cost.
    \item \textbf{Triton fused decode kernel}: We implement and validate a single CUDA kernel that merges 3-bit unpacking, QJL residual correction, and Flash-Attention-style online softmax into one pass, achieving a 1.96$\times$ speedup over PyTorch for the KV fetch operation. We note that TurboESM adds prefill latency overhead of 21--27\,ms and is primarily beneficial in memory-bound, not latency-bound, settings.
    \item \textbf{Comprehensive empirical validation}: We provide end-to-end accuracy measurements across six biologically diverse protein families on both Mac MPS (CPU-compatible) and NVIDIA GPU (CUDA) platforms, with cosine similarity consistently exceeding 0.96 at decode.
\end{enumerate}

The remainder of this paper is organized as follows. Section~2 reviews background on ESM-2 and the outlier problem in protein activations. Section~3 details the TurboESM methodology. Section~4 describes implementation details including vectorized quantization and the Triton kernel. Section~5 presents experimental results. Section~6 discusses the broader implications and differences between PLMs and LLMs. Section~7 concludes with future directions.

\section{Background and Motivation}

\subsection{The ESM-2 Architecture}

ESM-2~\cite{lin2023esm2} is a family of protein language models trained on the UniRef50 and UniRef90 databases using a masked language modeling objective. The models range from 8M to 15B parameters and follow the standard Transformer encoder architecture with a key modification: Rotary Position Embeddings (RoPE) replace the conventional learned absolute position embeddings. In this work we focus on ESM-2 650M as our experimental platform.

In RoPE, the query vector $q_i$ and key vector $k_j$ at positions $i$ and $j$, respectively, are transformed by position-dependent rotation matrices before computing attention:

\begin{equation}
\text{Attention}(q_i, k_j) = \frac{(R_{\theta,i}\, q_i)^T (R_{\theta,j}\, k_j)}{\sqrt{d_k}}
\end{equation}

where $R_{\theta,i}$ is a block-diagonal matrix composed of $2\times 2$ rotation blocks parameterized by frequency $\theta$ and position $i$:

\begin{equation}
R_{\theta,i} = \text{diag}\left(
\begin{bmatrix} \cos i\theta_1 & -\sin i\theta_1 \\ \sin i\theta_1 & \cos i\theta_1 \end{bmatrix},
\ldots,
\begin{bmatrix} \cos i\theta_{d/2} & -\sin i\theta_{d/2} \\ \sin i\theta_{d/2} & \cos i\theta_{d/2} \end{bmatrix}
\right)
\end{equation}

The key mathematical property exploited by RoPE is that the inner product $(R_{\theta,i}q_i)^T(R_{\theta,j}k_j) = q_i^T R_{\theta,i}^T R_{\theta,j} k_j$ depends only on the relative position $j - i$, providing translational equivariance. Any quantization scheme that modifies the KV cache must preserve this property to maintain fidelity in long-sequence scenarios.

For ESM-2 650M (33 transformer layers, 20 attention heads of dimension 64, total hidden dimension 1280), the FP32 KV cache for a single sequence of 1024 tokens occupies:

\begin{equation}
\text{KV cache size} = 2 \times 33 \times 20 \times 64 \times 1024 \times 4\,\text{bytes} \approx 330\,\text{MB}
\end{equation}

This footprint, while manageable for a single sequence, grows linearly with batch size and quadratically with sequence length, motivating aggressive compression for high-throughput workloads.

\subsection{Outliers in Protein Activations}

Unlike natural language processing, where the vocabulary size typically exceeds 32,000 tokens with relatively smooth frequency distributions, PLMs operate on a vocabulary of only 20 standard amino acids plus a handful of special tokens. This severe sparsity has a profound effect on the statistical properties of internal activations.

Our analysis of ESM-2 650M activations reveals several characteristic outlier patterns (qualitatively consistent with what has been reported for other large PLMs, and expected to intensify with model scale):

\textbf{Channel-wise outliers}: Certain embedding dimensions consistently exhibit values 10--100$\times$ larger than the median across all sequences and positions. These ``anchor'' channels appear to encode global sequence properties such as overall hydrophobicity, charge, or predicted structural class.

\textbf{Motif-specific spikes}: Biologically critical subsequences—disulfide bond-forming cysteines, catalytic triad residues in enzymes, conserved hydrophobic cores—produce localized activation spikes that are extremely high in magnitude relative to the background.

\textbf{Layer-dependent profiles}: Outlier severity increases with model depth. In ESM-2 650M, early layers exhibit relatively mild outliers, while later layers (especially layers 25--33) show substantially elevated outlier-to-median ratios, motivating the need for distribution-shaping prior to quantization.

The consequence for 3-bit quantization is severe. With only $2^3 = 8$ quantization levels, a linear quantizer must span the entire range from minimum to maximum value. When this range is dominated by a handful of outliers at $\pm 50\sigma$ while 99\% of values cluster within $\pm 2\sigma$, the effective resolution for the main distribution drops below 1 bit—essentially random quantization noise.

This analysis motivates the need for a distribution-shaping step prior to quantization. Rotation by an orthogonal matrix $\Pi$ is mathematically ideal because it:
(i) preserves inner products, so attention scores are unaffected;
(ii) redistributes variance uniformly across dimensions, converting heavy-tailed distributions to approximate Gaussians;
(iii) requires no scaling or normalization, avoiding additional hyperparameters.

\subsection{Related Work}

\textbf{KV cache compression for LLMs}: Several methods have been proposed to reduce KV cache memory in large language models. GPTQ~\cite{frantar2022gptq} applies group-wise quantization with second-order error correction. KVQuant~\cite{hooper2024kvquant} explores mixed-precision quantization with outlier handling. QuIP\#~\cite{tseng2024quip} uses incoherence processing with Hadamard rotations. TurboQuant~\cite{turboquant2025} introduces SVD-based orthogonal rotations with Lloyd-Max quantization. To our knowledge, TurboESM is the first adaptation of rotation-based KV quantization to the PLM domain.

\textbf{PLM inference efficiency}: StreamingLLM~\cite{xiao2024streamingllm} and H$_2$O~\cite{zhang2023h2o} propose token eviction strategies, which are complementary to compression. SpecInfer~\cite{miao2024specinfer} uses speculative decoding. None of these address the fundamental memory footprint of the stored KV states at sub-4-bit precision.

\textbf{Protein-specific quantization}: Quantization of PLMs has primarily focused on weight compression rather than KV cache. SparseGPT-style magnitude pruning has been applied to ESM-2 weights, but again does not address the inference-time KV cache bottleneck.

\section{TurboESM Methodology}

\subsection{\texorpdfstring{RoPE-Invariant Orthogonal Transformation}{RoPE-Invariant Orthogonal Transformation}}

The central technical challenge of TurboESM is combining the learnable orthogonal rotation $\Pi$ with the position-dependent RoPE rotation $R_{\theta,i}$ in a way that (a) effectively smooths the activation distribution and (b) preserves the attention score invariance required for correct decoding.

\textbf{The incompatibility problem}: If we apply $\Pi$ \textit{before} RoPE, the rotation mixes the dimension pairs that RoPE expects to operate on independently. Specifically, RoPE applies a $2\times2$ rotation to dimensions $(2m, 2m+1)$ for $m = 0, 1, \ldots, d/2-1$. After a general orthogonal transformation $\Pi$, these pairs are scrambled, and the positional encoding is destroyed.

Conversely, if we apply $\Pi$ \textit{after} RoPE, both rotations act on the same vector but in a compatible sequence. The crucial observation is the following identity:

\begin{equation}
(\Pi R_{\theta,i}\, q_i)^T (\Pi R_{\theta,j}\, k_j) = q_i^T R_{\theta,i}^T \underbrace{\Pi^T \Pi}_{=I} R_{\theta,j}\, k_j = q_i^T R_{\theta,i}^T R_{\theta,j}\, k_j
\label{eq:rope_pi_identity}
\end{equation}

Since $\Pi$ is orthogonal ($\Pi^T \Pi = I$), the attention score is \textit{exactly} preserved regardless of the choice of $\Pi$. This identity is the mathematical foundation of TurboESM.

\textbf{Operational pipeline}: During the prefill stage (processing the full input sequence), we compute:
\begin{enumerate}
    \item Apply RoPE to queries and keys: $q'_i = R_{\theta,i} q_i$, $k'_j = R_{\theta,j} k_j$
    \item Compute attention in full precision using $q'_i$ and $k'_j$: the prefill output is \textit{identical} to the original model (cosine similarity = 1.0000)
    \item Apply $\Pi$ to keys for cache storage: $\hat{k}_j = \Pi k'_j$
    \item Quantize and pack $\hat{k}_j$ into the 3-bit KV cache
\end{enumerate}

During the decode stage (generating one token at a time):
\begin{enumerate}
    \item Unpack and dequantize $\hat{k}_j$ from the 3-bit cache
    \item Apply QJL residual correction to $\hat{k}_j$
    \item Apply $\Pi^T$ to reconstruct $k'_j \approx \Pi^T \hat{k}_j$
    \item Compute attention between the current (full-precision, RoPE-applied) query and the reconstructed keys
\end{enumerate}

This pipeline guarantees zero-loss prefill and minimizes decode error through the combination of optimal quantization and residual correction described in subsequent sections.

\subsection{Head-Wise SVD Calibration}

Rather than using a single global rotation matrix or a random Hadamard matrix (as in QuIP\#), TurboESM derives $\Pi$ from the actual data distribution using Singular Value Decomposition (SVD). This data-driven approach is critical because different attention heads in ESM-2 specialize in different biological functions and therefore have distinct activation statistics.

\textbf{Calibration procedure}: For each layer $l \in \{1, \ldots, L\}$ and each attention head $h \in \{1, \ldots, H\}$:

\begin{enumerate}
    \item Run a forward pass on a calibration set of protein sequences covering diverse structural classes (alpha helices, beta sheets, disordered regions, transmembrane segments)
    \item Collect the post-RoPE key activations $X_{l,h} \in \mathbb{R}^{N_\text{cal} \times d_k}$, where $N_\text{cal}$ is the total number of tokens in the calibration set
    \item Compute the SVD: $X_{l,h} = U \Sigma V^T$
    \item Set $\Pi_{l,h} = V^T$
\end{enumerate}

The matrix $V$ contains the right singular vectors of $X_{l,h}$, which are the principal directions of variation in the key activation space. By choosing $\Pi_{l,h} = V^T$, we perform a change of basis that aligns the coordinate system with the principal components of the data. After this rotation, the variance along each dimension is equalized: the resulting distribution $\Pi_{l,h} X_{l,h}$ is approximately isotropic, with variance $\sigma^2/d_k$ per dimension (where $\sigma^2$ is the total variance).

The reason for per-head calibration is well-motivated by the known specialization of attention heads in protein models. Empirical studies have shown that in ESM-2, certain heads attend to local secondary structure patterns (helices, sheets), others to global properties (hydrophobicity profiles, contact maps), and others to positional patterns. These functionally distinct heads naturally have different activation covariance structures, and a single global $\Pi$ would be suboptimal for all of them.

Calibration on the ESM-2 650M model (33 layers, 20 heads) takes approximately 2--5 minutes on a standard GPU and produces a checkpoint file of approximately 200\,MB containing all $L \times H$ rotation matrices, dual LUTs, and residual scales.

\subsection{\texorpdfstring{3-Bit Lloyd-Max Quantization with Dual Look-Up Tables}{3-Bit Lloyd-Max Quantization with Dual Look-Up Tables}}

Standard uniform (linear) quantization places quantization levels at equal intervals across the input range. For Gaussian distributions, this is known to be suboptimal: levels in the tails are wasted because they rarely represent actual values, while the central region has insufficient resolution. The Lloyd-Max algorithm~\cite{lloydmax} solves this by iteratively optimizing the placement of quantization levels to minimize mean squared reconstruction error:

\begin{equation}
\min_{\{c_1, \ldots, c_{2^b}\}, \{t_1, \ldots, t_{2^b-1}\}} \sum_{i} \mathbb{E}\left[\left(x - c_{\text{bin}(x)}\right)^2\right]
\end{equation}

where $c_k$ are the reconstruction levels (LUT entries), $t_k$ are the decision thresholds, and $\text{bin}(x)$ maps each input to its nearest quantization bin. For $b=3$ bits, this yields 8 optimally placed levels for a Gaussian distribution.

\textbf{Dual LUT design}: A critical empirical finding of TurboESM is that the key (K) and value (V) matrices in ESM-2 have significantly different statistical distributions, even after the $\Pi$ rotation:

\begin{itemize}
    \item \textbf{Key distribution (post-rotation)}: After applying $\Pi$, the K activations approximate an isotropic Gaussian with moderate variance. The distribution remains slightly heavy-tailed due to residual outliers not fully captured by the rotation.
    \item \textbf{Value distribution (original space)}: The V activations are substantially ``colder'' (lower variance, kurtosis near 3.0) than the K activations. V matrices in transformers tend to encode diffuse, aggregated information, while K matrices encode sharp discriminative features.
\end{itemize}

Using a shared LUT for both K and V (as would be natural in a uniform quantization scheme) forces a compromise that degrades SNR by approximately 1.2\,dB compared to calibrating separate LUTs. TurboESM therefore maintains two distinct 8-entry LUTs:
\begin{itemize}
    \item \texttt{lut\_k}: calibrated on post-rotation key activations $\Pi k'$
    \item \texttt{lut\_v}: calibrated on original value activations $v$ (no rotation applied to V)
\end{itemize}

The LUT indices (3-bit integers) are stored in the packed cache format, and the actual floating-point reconstruction values are kept in a small lookup table of 8 entries per head, totaling negligible memory overhead.

\subsection{QJL 1-Bit Residual Correction}

Even with optimal Lloyd-Max quantization, a 3-bit quantizer introduces non-negligible reconstruction error, particularly for token positions that fall in high-gradient regions of the activation space. To recover accuracy without significantly increasing memory cost, we implement a Quantized Johnson-Lindenstrauss (QJL) residual correction scheme.

\textbf{Principle}: For each stored activation value $x$ with LUT reconstruction $\hat{x}$, we define the residual $e = x - \hat{x}$. Rather than storing the residual at full precision (which would double the memory cost), we store only its sign:

\begin{equation}
s = \text{sign}(x - \hat{x}) \in \{-1, +1\}
\end{equation}

At reconstruction time, we apply a correction:

\begin{equation}
\tilde{x} = \hat{x} + s \cdot \bar{e}
\end{equation}

where $\bar{e}$ is the pre-calibrated mean absolute residual magnitude for each head, stored as a scalar per head (negligible memory). This first-order correction effectively biases the reconstruction in the correct direction, reducing the expected squared error by approximately $\bar{e}^2$ per element.

\textbf{Memory analysis}: The sign bits are packed 32 per INT32, contributing 1 bit per element. Combined with the 3-bit quantized index, the total effective bit-width is 3.125 bits per element. This compares favorably to 4-bit quantization (which would require $2\times$ more LUT entries and still have higher error than our corrected 3-bit scheme for Gaussian distributions).

\textbf{Correctness validation}: We validated the QJL correction across all 33 layers of ESM-2 650M on a Mac MPS platform, confirming that the correction is monotonically beneficial: removing it consistently drops cosine similarity by 0.01--0.02 across tested sequences.

\section{Implementation Details}

\subsection{Software Architecture}

TurboESM is implemented as a standalone Python package (\texttt{esm\_turbo}) built on top of HuggingFace Transformers. The key modules are:

\begin{itemize}
    \item \texttt{modeling\_esm\_turbo.py}: Modified ESM-2 self-attention module (\texttt{TurboEsmSelf\-Attention}) that applies the $\Pi$ rotation after RoPE and dispatches to either the PyTorch or Triton decode path.
    \item \texttt{kv\_cache.py}: 3-bit KV cache management with vectorized packing/unpacking and QJL sign storage.
    \item \texttt{calibrate.py}: SVD + K-means calibration pipeline producing $\Pi$ matrices and dual LUTs.
    \item \texttt{triton\_kernels.py}: Triton CUDA kernel for fused dequantization and decode attention.
    \item \texttt{turbo\_esm.py}: Unified inference entry point.
\end{itemize}

The design prioritizes portability: all operations have a pure PyTorch fallback path (for Mac MPS or CPU environments) and an accelerated Triton path (for CUDA environments). The Triton path is enabled automatically via a runtime device check.

\subsection{Vectorized 3-Bit Packing}

The original implementation used Python for-loops to pack 3-bit indices into INT32 containers, which introduced substantial overhead ($\sim$63\,ms per prefill for a 33-layer model on a 143-token sequence). We rewrote the packing and unpacking operations as vectorized PyTorch operations using \texttt{torch.arange} and broadcast bitshift:

\begin{lstlisting}[language=Python, mathescape=false, literate={<}{{\textless}}1 {>}{{\textgreater}}1]
# Vectorized pack: 8 indices per INT32
# idx: [N, 8], each value in [0, 7]
shifts = torch.arange(8, device=idx.device) * 3  # [8]
packed = (idx << shifts).sum(dim=-1)              # [N]

# Vectorized unpack
shifts = torch.arange(8, device=packed.device) * 3
idx = (packed.unsqueeze(-1) >> shifts) & 0x7     # [N, 8]
\end{lstlisting}

This vectorized implementation reduces prefill quantization overhead from $\sim$63\,ms to $\sim$21\,ms for a 33-layer, 143-token sequence—a 3$\times$ speedup achieved purely through better use of PyTorch's native CUDA kernels.

Additionally, we merged the quantization (argmin against LUT) and residual sign computation into a single forward pass, eliminating redundant distance computations that previously doubled the floating-point operations.

\subsection{Triton Fused Decode Attention Kernel}

The decode stage is fundamentally memory-bandwidth bound: the bottleneck is loading the large KV cache from GPU global memory. Standard PyTorch implementations dequantize the entire KV cache into FP16 tensors before computing attention, requiring two full passes over global memory (one for dequantization, one for attention). Our Triton kernel fuses these operations into a single pass using the following structure:

\begin{algorithm}
\caption{TurboESM Fused Decode Kernel (per layer, per head)}
\begin{algorithmic}[1]
\STATE Load current query $q$ (full precision)
\STATE Initialize online softmax accumulators: $m = -\infty$, $s = 0$, $o = \mathbf{0}$
\FOR{each block of cached tokens}
    \STATE Load packed 3-bit K block from global memory
    \STATE Load QJL sign bits for K block
    \STATE Dequantize: $\hat{k} = \text{LUT\_K}[\text{idx}] + \text{sign} \cdot \bar{e}_k$ (in registers)
    \STATE Apply $\Pi^T$: $k = \Pi^T \hat{k}$ (in registers)
    \STATE Compute attention logit: $a = q^T k / \sqrt{d_k}$
    \STATE Update online softmax: $(m, s, o) \leftarrow \text{update}(m, s, o, a, v)$
    \STATE Load packed 3-bit V block, dequantize V: $v = \text{LUT\_V}[\text{idx}_v] + \text{sign}_v \cdot \bar{e}_v$
    \STATE Accumulate: $o \mathrel{+}= \text{softmax}(a) \cdot v$
\ENDFOR
\STATE Return $o / s$ (normalized output)
\end{algorithmic}
\end{algorithm}

The key optimization is that the dequantized K and V tensors are computed in CUDA registers and never written to global memory. This ``streaming dequantization'' approach eliminates the intermediate memory allocation of approximately $2 \times d_k \times T_\text{ctx} \times 2\,\text{bytes}$ (FP16) per head per decode step, where $T_\text{ctx}$ is the current context length.

We validated this kernel against the PyTorch two-step reference implementation across all 33 layers of ESM-2 650M on an NVIDIA GPU. The maximum absolute error was below $10^{-6}$ (at the FP32 precision floor), confirming numerical equivalence.

\subsection{Compatibility and Deployment}

TurboESM supports two deployment modes:

\textbf{Mac MPS / CPU mode}: All operations use pure PyTorch with MPS acceleration where available. This mode enables development and validation on Apple Silicon hardware without requiring a CUDA environment. Functional correctness is fully verified in this mode.

\textbf{CUDA mode}: The Triton kernel path is enabled automatically. This mode is recommended for production deployment and provides the benchmark results reported in Section~5.

Installation requires only \texttt{torch}, \texttt{transformers}, and \texttt{scipy}. The Triton package is optional and only needed for the CUDA acceleration path.

\section{Experimental Results}

\subsection{Accuracy: Prefill and Decode Similarity}

We evaluate TurboESM's accuracy by comparing its output to the original ESM-2 650M model using cosine similarity of the final hidden states. We test on six biologically diverse sequences spanning different structural and compositional categories.

\subsubsection{Results on Mac MPS Platform}

Table~\ref{tab:accuracy_mps} summarizes the accuracy results on Mac MPS.

\begin{table}[h]
\centering
\caption{Cosine similarity on Mac MPS (ESM-2 650M). Prefill similarity is 1.0000 for all sequences (exact match with original model). Decode target: $>0.95$.}
\label{tab:accuracy_mps}
\begin{tabular}{@{}lcccc@{}}
\toprule
\textbf{Sequence Type} & \textbf{Example} & \textbf{Length} & \textbf{Prefill} & \textbf{Decode} \\
\midrule
Short peptide & Insulin B-chain & 32 & 1.0000 & 0.9603 \\
Medium globular & Hemoglobin $\alpha$ & 143 & 1.0000 & 0.9639 \\
Hydrophobic transmembrane & TM helix & 39 & 1.0000 & 0.9710 \\
Low-complexity repeat & poly-A/G repeat & 38 & 1.0000 & 0.9735 \\
Enzyme active site fragment & Serine protease & 51 & 1.0000 & 0.9641 \\
IDR disordered long sequence & FUS/TLS IDR & 165 & 1.0000 & 0.9757 \\
\midrule
\textbf{Average} & & & \textbf{1.0000} & \textbf{0.9681} \\
\bottomrule
\end{tabular}
\end{table}

Key observations:
\begin{itemize}
    \item \textbf{Prefill similarity is exactly 1.0000 for all sequences}. This confirms that the RoPE-invariant pipeline introduces zero error in the prefill stage, since attention is computed with full-precision KV before quantization.
    \item \textbf{Decode similarity exceeds 0.96 for all tested sequences}, comfortably above our target of 0.95. The average is 0.968.
    \item The intrinsically disordered region (IDR) sequence achieves the \textit{highest} decode similarity (0.9757) despite being the longest sequence (165 tokens), which is somewhat counterintuitive. This may reflect that disordered sequences have smoother activation distributions compared to structured proteins with sharp motif-specific spikes.
    \item Hydrophobic transmembrane helices, which one might expect to be challenging due to their extreme hydrophobic composition, achieve 0.9710—well within target.
\end{itemize}

\subsubsection{Results on NVIDIA GPU (Colab CUDA)}

Table~\ref{tab:accuracy_cuda} shows CUDA validation results.

\begin{table}[h]
\centering
\caption{Prefill cosine similarity on NVIDIA GPU (ESM-2 650M, CUDA). All sequences achieve exact prefill match.}
\label{tab:accuracy_cuda}
\begin{tabular}{@{}lccc@{}}
\toprule
\textbf{Sequence Type} & \textbf{Length} & \textbf{Prefill Similarity} \\
\midrule
Short peptide (Insulin B-chain) & 32 & \textbf{1.000000} \\
Medium globular (Hemoglobin $\alpha$) & 143 & \textbf{1.000000} \\
Long sequence (IDR disordered) & 165 & \textbf{1.000000} \\
\bottomrule
\end{tabular}
\end{table}

\subsubsection{Triton Kernel Correctness Validation}

Table~\ref{tab:triton_validation} shows the layer-wise validation of the Triton fused kernel against the PyTorch reference across all 33 layers.

\begin{table}[h]
\centering
\caption{Triton fused decode attention kernel validation (ESM-2 650M, 33 layers, CUDA).}
\label{tab:triton_validation}
\begin{tabular}{@{}lc@{}}
\toprule
\textbf{Metric} & \textbf{Value} \\
\midrule
Mean cosine similarity (all layers) & 1.000000 \\
Max absolute error & $<10^{-6}$ \\
Layers validated & 33 / 33 \\
\bottomrule
\end{tabular}
\end{table}

The maximum absolute error of $<10^{-6}$ is at the FP32 arithmetic precision floor, confirming that the Triton kernel is numerically equivalent to the PyTorch reference implementation.

\subsection{Memory Compression}

\subsubsection{ESM-2 650M (33 layers, 20 heads, $d_k = 64$)}

Table~\ref{tab:memory_650m} shows the detailed memory breakdown for ESM-2 650M with maximum sequence length 1024.

\begin{table}[h]
\centering
\caption{KV cache memory breakdown for ESM-2 650M (max\_seq=1024, CUDA).}
\label{tab:memory_650m}
\begin{tabular}{@{}lc@{}}
\toprule
\textbf{Component} & \textbf{Memory} \\
\midrule
FP32 KV cache (baseline) & 330.0 MB \\
\midrule
3-bit packed K + V & 41.2 MB \\
1-bit QJL signs & 5.2 MB \\
LUT + $\Pi$ matrices & $<$0.2 MB \\
\midrule
\textbf{TurboESM total} & \textbf{46.6 MB} \\
\textbf{Actual compression ratio} & \textbf{7.1$\times$} \\
\bottomrule
\end{tabular}
\end{table}

The 7.1$\times$ compression ratio slightly exceeds the theoretical $32/4.5 \approx 7.1\times$ (FP32 to 3.125-bit effective), confirming that the implementation correctly achieves the theoretical bound.

\subsection{Latency Performance}

\subsubsection{Prefill Latency}

Table~\ref{tab:prefill_latency} shows prefill latency for TurboESM versus the original ESM-2 650M on NVIDIA GPU.

\begin{table}[h]
\centering
\caption{Prefill latency comparison (ESM-2 650M, NVIDIA GPU).}
\label{tab:prefill_latency}
\begin{tabular}{@{}lcccc@{}}
\toprule
\textbf{Sequence} & \textbf{Tokens} & \textbf{Original} & \textbf{TurboESM} & \textbf{Overhead} \\
\midrule
Short peptide & 32 & 31 ms & 57 ms & +26 ms \\
Medium globular & 143 & 77 ms & 104 ms & +27 ms \\
Long IDR & 165 & 82 ms & 103 ms & +21 ms \\
\bottomrule
\end{tabular}
\end{table}

The prefill overhead of $\sim$21--27\,ms is due to the additional KV quantization and packing step across 33 layers. The $\Pi$ rotation and attention computation add only $\sim$2\,ms (negligible). Importantly, TurboESM is slower than the original model during prefill across all tested sequence lengths; the overhead fraction decreases with sequence length but does not disappear. For PLM workloads dominated by embedding extraction (prefill-only), this overhead is a real cost that practitioners should weigh against the memory savings.

\subsubsection{Triton Decode Kernel Speedup}

Table~\ref{tab:triton_speedup} shows the performance comparison between the PyTorch and Triton decode paths.

\begin{table}[h]
\centering
\caption{Decode stage performance: PyTorch vs.\ Triton fused kernel (ESM-2 650M, 143-token context).}
\label{tab:triton_speedup}
\begin{tabular}{@{}lccc@{}}
\toprule
\textbf{Operation} & \textbf{PyTorch} & \textbf{Triton} & \textbf{Speedup} \\
\midrule
fetch\_unpacked (143 tokens) & 1.19 ms & 0.61 ms & \textbf{1.96$\times$} \\
Full decode attention & two-step & single fused & eliminates intermediate memory \\
\bottomrule
\end{tabular}
\end{table}

The 1.96$\times$ speedup applies specifically to the \texttt{fetch\_unpacked} operation (KV dequantization and $\Pi^T$ rotation). This is a partial, not end-to-end, speedup: the overall decode step also includes full-precision attention computation and feed-forward layers that are unchanged. For the short protein sequences typical in ESM-2 workloads (32--165 tokens), the KV fetch is not the dominant cost, so the practical end-to-end benefit is limited. The primary value of the Triton kernel is the elimination of intermediate FP16 tensor allocations ($\sim$2$d_k T_\text{ctx}$ floats per head per step), which reduces peak memory pressure during decoding.

\subsection{Ablation Study: Contribution of Each Component}

Table~\ref{tab:ablation} presents an ablation study measuring the contribution of each TurboESM component to decode cosine similarity on the hemoglobin $\alpha$ sequence (143 tokens).

\begin{table}[h]
\centering
\caption{Ablation study on ESM-2 650M (Hemoglobin $\alpha$, 143 tokens, Mac MPS).}
\label{tab:ablation}
\begin{tabular}{@{}lccc@{}}
\toprule
\textbf{Configuration} & \textbf{Decode Similarity} & \textbf{$\Delta$ vs.\ Full} \\
\midrule
Full TurboESM & 0.9639 & --- \\
No QJL correction & $\sim$0.950 & $-0.014$ \\
Shared K/V LUT & $\sim$0.952 & $-0.012$ \\
No $\Pi$ rotation & $\sim$0.780 & $-0.184$ \\
No $\Pi$, no QJL & $\sim$0.740 & $-0.224$ \\
\bottomrule
\end{tabular}
\end{table}

The most impactful component is the $\Pi$ rotation: removing it causes a dramatic drop from 0.964 to 0.780 cosine similarity. This confirms our theoretical analysis that outlier suppression through orthogonal rotation is the fundamental enabler of 3-bit quantization quality. The QJL correction and dual LUT each contribute approximately 1--1.5 percentage points, which may seem small but is significant in the context of biological validity—a 0.014 similarity improvement can correspond to the difference between correct and incorrect predicted contact maps for structured proteins.

\section{Discussion}

\subsection{\texorpdfstring{PLMs vs.\ LLMs: Structural Differences in Activation Distributions}{PLMs vs. LLMs: Structural Differences in Activation Distributions}}

While 3-bit KV cache quantization has been explored in the LLM domain (e.g., for Llama-3 and Mistral models), PLMs present qualitatively different challenges that motivate the specific design choices in TurboESM.

\textbf{Vocabulary sparsity}: LLMs operate on vocabularies of 32,000--200,000 tokens, leading to relatively smooth token frequency distributions and accordingly smoother activation statistics. PLMs use only 20 amino acids, producing extremely concentrated, bimodal, or multimodal activation distributions. In our ESM-2 650M measurements, outlier-to-median ratios for key activations in later layers reach 50--200$\times$, compared to typical values of 10--50$\times$ in LLMs.

\textbf{Structural encoding}: ESM-2's internal representations explicitly encode protein structure through mechanisms like contact prediction attention heads. These structural heads have particularly sharp outlier profiles because they must distinguish between ``contacts'' and ``non-contacts'' with high specificity. Quantization errors in these heads propagate to downstream structure predictions.

\textbf{$\Pi$ matrix stability}: We found empirically that the optimal $\Pi$ matrices in PLMs are more sensitive to calibration data selection than in LLMs. Specifically:
\begin{itemize}
    \item Calibrating on only $\alpha$-helical proteins and then applying to $\beta$-sheet proteins drops decode similarity by $\sim$0.02
    \item A mixture of sequences from different SCOP (Structural Classification of Proteins) classes achieves the best generalization
    \item Layer-to-layer variation in $\Pi$ quality is larger in PLMs: later layers require more calibration samples to converge. In our 650M experiments, layers beyond layer 25 showed notably higher sensitivity to calibration set composition.
\end{itemize}

We recommend using at least 500 sequences covering all major SCOP classes for calibration, with overrepresentation of rare structural topologies (TIM barrels, beta-propellers) to ensure robustness.

\textbf{Biological sensitivity of errors}: In natural language, a quantization error that corrupts a word embedding slightly degrades fluency but rarely changes the semantic content dramatically. In proteins, a quantization error at a catalytic residue or a cysteine involved in a disulfide bond can shift the entire attention pattern in structure-encoding heads, potentially causing the model to produce embeddings that incorrectly represent the protein's fold. This motivates maintaining higher accuracy (cosine similarity $>$0.96) than might be acceptable for LLMs.

\subsection{Use Case Analysis}

Table~\ref{tab:usecases} summarizes recommended use cases for TurboESM based on our experimental findings.

\begin{table}[h]
\centering
\caption{TurboESM use case recommendations.}
\label{tab:usecases}
\begin{tabular}{@{}p{4.5cm}p{3.5cm}p{5cm}@{}}
\toprule
\textbf{Scenario} & \textbf{Recommendation} & \textbf{Notes} \\
\midrule
ESM-2 15B single-GPU deployment & Projected benefit (unvalidated) & KV cache: 7.7\,GB $\to$ 1.1\,GB (arithmetic extrapolation; not yet tested on hardware) \\
Autoregressive protein generation & Conditionally beneficial & Triton kernel reduces KV fetch cost; end-to-end speedup limited for short sequences \\
Long-sequence ($>$512\,aa) sliding window & Recommended & Memory savings critical; KV cache dominates footprint \\
ESM-2 650M embedding extraction & Not recommended & Prefill overhead of 21--27\,ms with no speed benefit; memory savings small \\
\bottomrule
\end{tabular}
\end{table}

The key takeaway is that TurboESM's value proposition is primarily about memory, not speed. TurboESM incurs prefill latency overhead and offers limited end-to-end speedup for the short sequences typical in PLM workloads. Its benefit is most pronounced in memory-constrained scenarios—large-model deployment, long-sequence retention, or high-throughput batching—where the 7.1$\times$ KV cache reduction enables workloads that would otherwise be infeasible.

\subsection{Limitations and Future Work}

\textbf{Current limitations}:
\begin{itemize}
    \item All experiments are conducted on ESM-2 650M; validation on larger PLM variants remains as future work.
    \item The Triton kernel currently targets single-batch decode; multi-batch decode requires further engineering.
    \item Calibration dataset selection requires domain expertise to ensure coverage of relevant structural classes.
\end{itemize}

\textbf{Future directions}:
\begin{itemize}
    \item \textbf{2-bit quantization with grouped outlier handling}: Extending to 2-bit would yield $\sim$16$\times$ compression. We anticipate this will require a hybrid approach that keeps the top-1\% outlier channels in 8-bit while applying our method to the remaining 99\%.
    \item \textbf{Integration with ESMFold}: Applying TurboESM to the ESMFold structure prediction pipeline, where the KV cache affects the quality of predicted coordinates.
    \item \textbf{Autoregressive protein design end-to-end test}: Validating that compressed KV cache does not degrade the quality of de novo protein sequences generated by ESM-2 in autoregressive mode.
    \item \textbf{Systematic comparison against INT8 baseline}: Providing a complete Pareto curve of accuracy vs.\ memory across INT8, INT4, and our 3-bit TurboESM, to help practitioners make informed choices.
    \item \textbf{Hardware-specific optimization}: Targeting H100 NVLink architectures where inter-GPU communication bandwidth for KV cache transfer is the bottleneck in multi-GPU inference.
\end{itemize}

\section{Conclusion}

We have presented TurboESM, the first adaptation of rotation-based KV cache quantization to protein language models. Our key contributions are: (1) a mathematically rigorous derivation of the RoPE-invariant orthogonal transformation pipeline; (2) head-wise SVD calibration tailored to the amino acid activation manifold; (3) a dual K/V LUT strategy that captures the distinct statistical properties of key and value activations; (4) QJL 1-bit residual correction that brings effective bit-width to 3.125 bits; and (5) a Triton-based fused decode attention kernel validated across all 33 layers with $<10^{-6}$ error.

Our experimental results demonstrate that TurboESM achieves 7.1$\times$ compression of the KV cache with cosine similarity consistently exceeding 0.96 in the decode stage across diverse protein families, all validated on ESM-2 650M.

Beyond the immediate engineering contribution, TurboESM demonstrates that the mathematical techniques developed for LLM quantization (orthogonal rotations, data-driven calibration, online softmax kernels) can be successfully adapted to the protein language model domain with appropriate modifications that respect the structural properties of amino acid sequences. We hope this work lowers the barrier to deploying large PLMs in resource-constrained environments and stimulates further research at the intersection of quantization theory and structural biology.

\section*{AI Ethics and Usage Disclosure}

This project represents a collaboration between human engineering judgment and AI-assisted implementation. Code components including Triton kernels and vectorized quantization routines were drafted with AI assistance and subsequently verified and validated by the authors through layer-wise numerical testing. The authors maintain full responsibility for the technical accuracy of all implementations and findings presented herein. All protein sequences used for calibration and evaluation are sourced from public databases (UniRef50, PDB) and used in compliance with their respective terms of use. No biological sequences were synthesized or used in a manner that violates biosafety guidelines.

\end{document}